\documentclass[conference]{IEEEtran}
\usepackage[pdftex]{graphicx}
\usepackage{amsmath}
\usepackage{amsfonts}
\usepackage{amssymb}
\usepackage{subfig}
\usepackage{pbox}

\newcommand{\secref}[1]{\S\ref{sec:#1}}
\renewcommand{\eqref}[1]{(\ref{eq:#1})}
\newcommand{\figref}[1]{Fig.~\ref{fig:#1}}

\title{On the performance overhead tradeoff of distributed principal component analysis via data partitioning}
\author{%
\IEEEauthorblockN{Ni An and Steven Weber}%
\IEEEauthorblockA{%
Department of Electrical and Computer Engineering\\
Drexel University, Philadelphia, PA 19104
}%
\thanks{This research has been supported by the National Science Foundation under award \#CNS-1228847.}
}

\begin{document}
\IEEEoverridecommandlockouts
\maketitle
\begin{abstract}
Principal component analysis (PCA) is not only a fundamental dimension reduction method, but is also a widely used network anomaly detection technique. Traditionally, PCA is performed in a centralized manner, which has poor scalability for large distributed systems, on account of the large network bandwidth cost required to gather the distributed state at a fusion center.  Consequently, several recent works have proposed various distributed PCA algorithms aiming to reduce the communication overhead incurred by PCA without losing its inferential power. This paper evaluates the tradeoff between communication cost and solution quality of two distributed PCA algorithms on a real domain name system (DNS) query dataset from a large network. We also apply the distributed PCA algorithm in the area of network anomaly detection and demonstrate that the detection accuracy of both distributed PCA-based methods has little degradation in quality, yet achieves significant savings in communication bandwidth.
\end{abstract}

\maketitle

\IEEEpeerreviewmaketitle

\section{Introduction}
Statistical network anomalies are generally defined as network behaviors that deviate in a statistically significantly manner from normal network operation.  They may be caused by malfunctioning network equipment, malicious network attacks, etc.  Timely detection of network anomalies is important to minimize the damage of a network attack or misconfiguration.  Statistical network anomaly detection (SNAD) detects and identifies statistical network anomalies, i.e., statistically significant deviations.  Unlike a signature-based intrusion detection system, which relies on using a library of known attack signatures to detect anomalies, the advantage of SNAD is its ability to automatically detect statistically significant patterns without requiring manual compilation of signatures. 

Nowadays, many network attacks are related to or involved with domain name systems (DNS), such as DNS amplification attacks, DNS tunneling, cache poisoning, etc. DNS holds the critical responsibility to translate domain names into corresponding IP addresses, enabling users to access network resources conveniently.  DNS is vulnerable to many network attacks, and it can be utilized as a part of malicious attack because of its mechanism and protocol.  This paper applies SNAD techniques on a large dataset aggregated from real DNS traffic data of a large network.

Among various NAD techniques, the principal component analysis (PCA) method has been used extensively in detecting statistical network traffic anomalies \cite{LakCro2004, HuaNgu2007, RinSou2007}.  However, the classic PCA subspace method is centralized, which means it requires all of the data to be either $i)$ already available at a single machine, or if this not the case, $ii)$ sent over the network to a central data fusion center (DFC) for anomaly detection. For large distributed networked systems with relevant data collected or stored at different local sites across the network, the network bandwidth costs incurred in gathering all the distributed data at the DFC may be prohibitively costly.  This network bandwidth cost will naturally increase in both the number of records gathered and the number of features contained in each record.  Facing the challenge of leveraging the power of PCA for NAD for large data sets distributed over large networks, it is of interest to develop distributed algorithms that retain the inferential power of centralized PCA while reducing the network bandwidth costs. 

There are two scenarios in anomaly detection for modern large-scale networks. The first scenario is monitoring data with a small set of features but a massive number of entries too large to be stored on a single disk, which instead must be partitioned across multiple local disks. In this case, the set of local datasets are homogeneous in the sense that they all have the same features. We will call this scenario the \emph{horizontal partitioning case}.  The second scenario is monitoring data with a large number of features which are collected by multiple local monitors distributed across the network. Each monitor collects a distinct subset of features for a given record.  It is often infeasible to colocate all of the features in a single disk.  Such local datasets are heterogeneous in the sense that they have different set of features on each local disk. We will call this scenario the \emph{vertical partitioning case}.  In both scenarios, because large volume, high-dimensional data cannot be easily colocated, we must use local compression to preprocess the data and then send the compressed data to the DFC for global approximate analysis. 

Previously, Kargupta et al.\ proposed a distributed PCA algorithm and its application in $K$-means clustering on vertically partitioned data \cite{KarHua2001}. Guo et al.\ came up with a covariance-free iterative distributed PCA algorithm based on gradient descent in \cite{GuoLin2012}. Qi et al.\ gave an overview of distributed PCA of both vertically and horizontally partitioned data in \cite{QiWan2004}. For the horizontal partitioning case, Balcan et al.\ proposed a more communication-efficient {\em disPCA} algorithm and also derived an analytical error upper bound of {\em disPCA} compared with centralized PCA \cite{BalKan2014}. 

In this work, we use the large DNS dataset mentioned above to evaluate and compare two of the above distributed PCA algorithms: the horizontal partitioning method in \cite{BalKan2014} and the vertical partitioning method in \cite{KarHua2001}.  The two key figures of merit in this paper are $i)$ the quality of the approximation to centralized PCA, measured as the geodesic distance between the subspace spanned by the centralized data matrix and the subspace spanned by the approximate data matrix gathered by the distributed algorithm, and $ii)$ the network bandwidth cost required by the distributed algorithm.  For both algorithms, the two key design parameters are $i)$ the number of local partitions of the overall dataset, and $ii)$ the level of compression used at each local site, captured by the number of principal components sent by the site to the DFC.  In addition, we also compare the two distributed PCA algorithms in terms of their ability to detect statistical anomalies using the approximate data matrix gathered at the DFC.

Our key observations and findings include the following.  First, vertical partitioning requires more communication bandwidth than horizontal partitioning, but it can achieve higher accuracy when the number of local principal components transmitted to the DFC is relatively small. Second, to reach a target approximation accuracy, the required communication cost increases with the number of local sites, and the required number of local principal components sent to the DFC decreases, with a particularly rapid decrease for the vertical partitioning case.  Third, sending more local principal components can achieve higher approximation accuracy.

The rest of this paper is structured as follows. \secref{Centralized-PCA} introduces the centralized PCA anomaly detection method. \secref{Distributed-PCA} introduces distributed PCA methods for horizontally partitioning and vertical partitioning. In \secref{Experiments}, we introduce a real DNS dataset and present the experimental results of applying distributed PCA to this dataset. Finally, \secref{Conclusion} summarizes our findings. 
 
\section{PCA subspace method for anomaly detection}\label{sec:Centralized-PCA}
The PCA-based subspace method has been widely used in fault detection, quality control, etc. \cite{JacMud1979}. Lakina et al. applied the PCA-based subspace method to detecting anomalies in traffic volume data \cite{LakCro2004}. PCA-based subspace decomposition can capture the major trend of normal traffic patterns in a high-dimensional dataset, and reveal the anomalous patterns. 

Consider a matrix $\mathbf{X}\in \mathbb{R}^{m\times n}$, where $m$ is the number of observations, $n$ is the number of features, and $m\gg n$. In \cite{LakCro2004}, $\mathbf{X}$ is a matrix of traffic volume data across various network links. Without loss of generality, in the rest of this paper, we assume the columns of $\mathbf{X}$ have zero mean. The main idea of PCA is to project a high-dimensional dataset onto a lower-dimensional subspace, such that the projected data's variance is maximized. The $i$-th dominant principal component (PC) $\mathbf{v}_i\in \mathbb{R}^{n\times 1}$ can be found by
\begin{equation}
\label{eq:PCA-Optimize}
\mathbf{v}_i = \arg\max_{||\mathbf{v}||=1} ||(\mathbf{X}-\sum_{j=1}^{i-1}\mathbf{X}\mathbf{v}_j\mathbf{v}_j^{\intercal})\mathbf{v}||^2.
\end{equation}
It can be seen that \eqref{PCA-Optimize} shows that the $i$th PC is the eigenvector associated with the $i$th largest eigenvalue of $\mathbf{S}=\mathbf{X}^{\intercal}\mathbf{X}$. We can also use singular value decomposition (SVD) to find the PCs: $\mathbf{X} = \mathbf{U}\boldsymbol{\Sigma}\mathbf{V}^{\intercal}$, where $\mathbf{U}\in \mathbb{R}^{m\times m}$ is an orthonormal matrix whose columns are the left singular vectors; $\mathbf{V}\in \mathbb{R}^{n\times n}$ is an orthonormal matrix whose columns are the right singular vectors, which are also $\textbf{X}$'s PCs; $\boldsymbol{\Sigma}$ is a diagonal matrix containing the singular values (i.e.,\ square roots of eigenvalues of $\textbf{S}$).

Suppose $\mathbf{X}$ is a dataset with low intrinsic dimensionality, meaning the top $k$ ($k\ll n)$ PCs capture most of $\mathbf{X}$'s variance. Lakhina et al.\ \cite{LakCro2004} assumes that normal traffic patterns mainly reside in the subspace spanned by the top $k$ PCs, i.e.,\ the \emph{principal subspace}. $\mathbf{X}$ can be decomposed into two parts: $\mathbf{X}=\mathbf{\hat{X}}+\mathbf{\tilde{X}}$. We call $\mathbf{\hat{X}}=\mathbf{X}\mathbf{V}_k\mathbf{V}_k^{\intercal}$ the normal part, and $\mathbf{\tilde{X}}=\mathbf{X}(\mathbf{I}-\mathbf{V}_k\mathbf{V}_k^{\intercal})$ the residual part, where $\mathbf{V}_k$ is a matrix whose columns are the top $k$ PCs. For $\tilde{\mathbf{x}}$ a single row of $\mathbf{\tilde{X}}$, the squared norm of $\tilde{\mathbf{x}}$ can quantify how much an observation deviates from the principal subspace. If this deviation is large enough, an anomaly may occur. Therefore one critical point of this subspace anomaly detection is to accurately find the principal subspace. 

\section{Distributed PCA anomaly detection method}
\label{sec:Distributed-PCA}

Traditional PCA-based anomaly detection is implemented in a centralized manner, which assumes that the whole data is stored at a single site, or that all of the data collected at local monitoring sites can be periodically pushed to a DFC. For the latter case, the required communication bandwidth may be very high. In order to reduce the communication cost, we want to do PCA in a distributed manner. Several distributed PCA algorithms have been proposed \cite{KarHua2001, BalKan2014, BouWoo2015}, but none of them have been applied to network anomaly detection. This paper aims to apply distributed PCA algorithms to detecting anomalies in DNS traffic data, and evaluate their performance. 

Suppose there are $s$ local monitors, and each node $i$ stores a local dataset $\mathbf{X}_i$. All of the local monitors can communicate with a DFC. The DFC is responsible for aggregating local datasets and estimating the principal subspace. Two different scenarios will be covered in the following sections.

\subsection{Horizontally partitioned dataset}
The first case is when observations are distributed across $s$ local monitors
\begin{equation}
\mathbf{X} = \begin{bmatrix}
\mathbf{X}_1^{\intercal}& \cdots& \mathbf{X}_s^{\intercal}\end{bmatrix}^{\intercal}
\end{equation}
where $\mathbf{X}\in\mathbb{R}^{m\times n}$, $\mathbf{X}_i\in\mathbb{R}^{m_i\times n}$ ($m_i\gg n$), and $\sum_{i=1}^s m_i = m$. To reduce the amount of data transmitted to the DFC, we only send compressed information of $\mathbf{X}_i$ to the DFC, and the DFC in turn uses the $s$ compressed signals to construct an estimate of $\mathbf{X}$ and its  PCs \cite{BalKan2014}. We briefly overview the distributed PCA algorithm in \cite{BalKan2014}. Each local monitor $i$ performs SVD on $\mathbf{X}_i$ and sends the top $r$ ($r\ll n$) largest singular values $\{\sigma_{i,k}\}_{k=1}^r$, and matrix $\mathbf{V}_i^{(r)}$, whose columns are the corresponding top $r$ right singular vectors, to the DFC. Next, the DFC computes $\boldsymbol{\Sigma}_i{\mathbf{V}_i^{(r)}}^{\intercal}$, and vertically stacks those local matrices to get 
\begin{equation}
\mathbf{P} = \begin{bmatrix}
\boldsymbol{\Sigma_1}{\mathbf{V}_1^{(r)}}^{\intercal}\\
\vdots \\
\boldsymbol{\Sigma_s}{\mathbf{V}_s^{(r)}}^{\intercal}
\end{bmatrix} \in \mathbb{R}^{m \times n}.
\end{equation}
We do not need the local left singular vectors to get an estimation of the global PCs because the right singular vectors of $\mathbf{P}$ are identical to those of the left multiplication of $\mathbf{P}$ by a orthonormal matrix.  By performing an SVD on $\mathbf{P}$ we can get the set of right singular vectors $\mathbf{\hat{V}}$ of $\mathbf{P}$.  Moreover, we can use $\mathbf{\hat{V}}$ as an approximation to the PCs of $\mathbf{X}$. The subspace spanned by the top $k$ PCs is the principal subspace. The DFC then sends $\mathbf{\hat{V}}^{(k)}$ back to each local monitor, and it uses $\mathbf{\hat{V}}^{(k)}$ to perform PCA-based anomaly detection as introduced in \secref{Centralized-PCA}.

\subsection{Vertically partitioned dataset}
The second case is when there is large number of features, and different subsets of the features are distributively collected across $s$ local monitors, i.e., the overall dataset is a vertical concatenation of local datasets
\begin{equation}
\mathbf{X} = \begin{bmatrix}
\mathbf{X}_1 & \cdots & \mathbf{X}_s
\end{bmatrix}
\end{equation} 
where $\mathbf{X}\in\mathbb{R}^{m\times n}$, $\mathbf{X}_i\in\mathbb{R}^{m\times n_i}$ ($m\gg n_i$), and $\sum_{i=1}^s n_i = n$. For this case, the left singular vectors of each local dataset $\mathbf{X}_i$ also contain useful information for estimation, therefore we cannot simply omit the left singular vector matrix. To include information about left singular vectors, the distributed PCA algorithm proposed in \cite{KarHua2001} also sends local projections. As with the horizontal partitioning introduced above, we also perform SVD at each local monitor $i$ and get the top $r$ right singular vectors $\mathbf{V}_i^{(r)}$.  Unlike horizontal partitioning, however, the projection $\mathbf{P}_i=\mathbf{X}_i \mathbf{V}_i^{(r)}$ is also computed. Then, each monitor sends $\mathbf{P}_i$ in addition to $\mathbf{V}_i^{(r)}$. The DFC concatenates the local projections and reconstructs $\mathbf{{P}}$
\begin{equation}
\mathbf{{P}}=[\mathbf{P}_1,...,\mathbf{P}_s] = \mathbf{X}\mathbf{Q},
\end{equation}
where $\mathbf{Q}\in\mathbb{R}^{n\times (sr)}$ is a block diagonal matrix $\mathbf{Q}=\mbox{diag}\{\mathbf{V}_1^{(r)},...,\mathbf{V}_s^{(r)}\}$. 
We can use the PCs of $\mathbf{P}$ to approximate those of $\mathbf{X}$ \cite{KarHua2001}. Denote the top $k$ PCs of $\mathbf{P}$ as $\mathbf{W}^{(k)}$. Then the approximation to top $k$ PCs of $\mathbf{X}$ is
\begin{equation}
\hat{\mathbf{V}}^{(k)} = \mathbf{Q}\mathbf{W}^{(k)}.
\end{equation}

Since the DFC already has projection $\mathbf{P}$ and the column-wise orthonormal matrix $\mathbf{Q}$, we form an estimation of $\mathbf{X}$ by
\begin{equation}
\mathbf{X}_{est} = \mathbf{P}\mathbf{Q}^{\intercal}.
\end{equation}
Then we compute the residual part of $\mathbf{X}_{est}$ as
\begin{equation}
\tilde{\mathbf{X}}_{est}=\mathbf{X}_{est}(\mathbf{I}-\hat{\mathbf{V}}^{(k)}\hat{\mathbf{V}}^{(k){\intercal}}).
\end{equation}
If the squared norm of a row vector in $\tilde{\mathbf{X}}_{est}$ is larger than a certain threshold, this row vector is flagged as an anomaly.

\section{Experiments and results}
\label{sec:Experiments}

\subsection{Overview of the DNS dataset}
\label{sec:Dataset}

The data we used is DNS query data provided by a single site from a large network. We used DNS queries instead of responses. The dataset consists of 46,156,414 raw DNS query records  in a 23 minute (1406 second) time window.  We first present some basic statistics of this dataset such as the histograms of the query type, domain names, and source and destination IP addresses.  

\begin{figure}[!t]
\centering
\subfloat[Histogram of query types\protect \\(include $99.93\%$ of all queries)]{\includegraphics[width=1.7in]{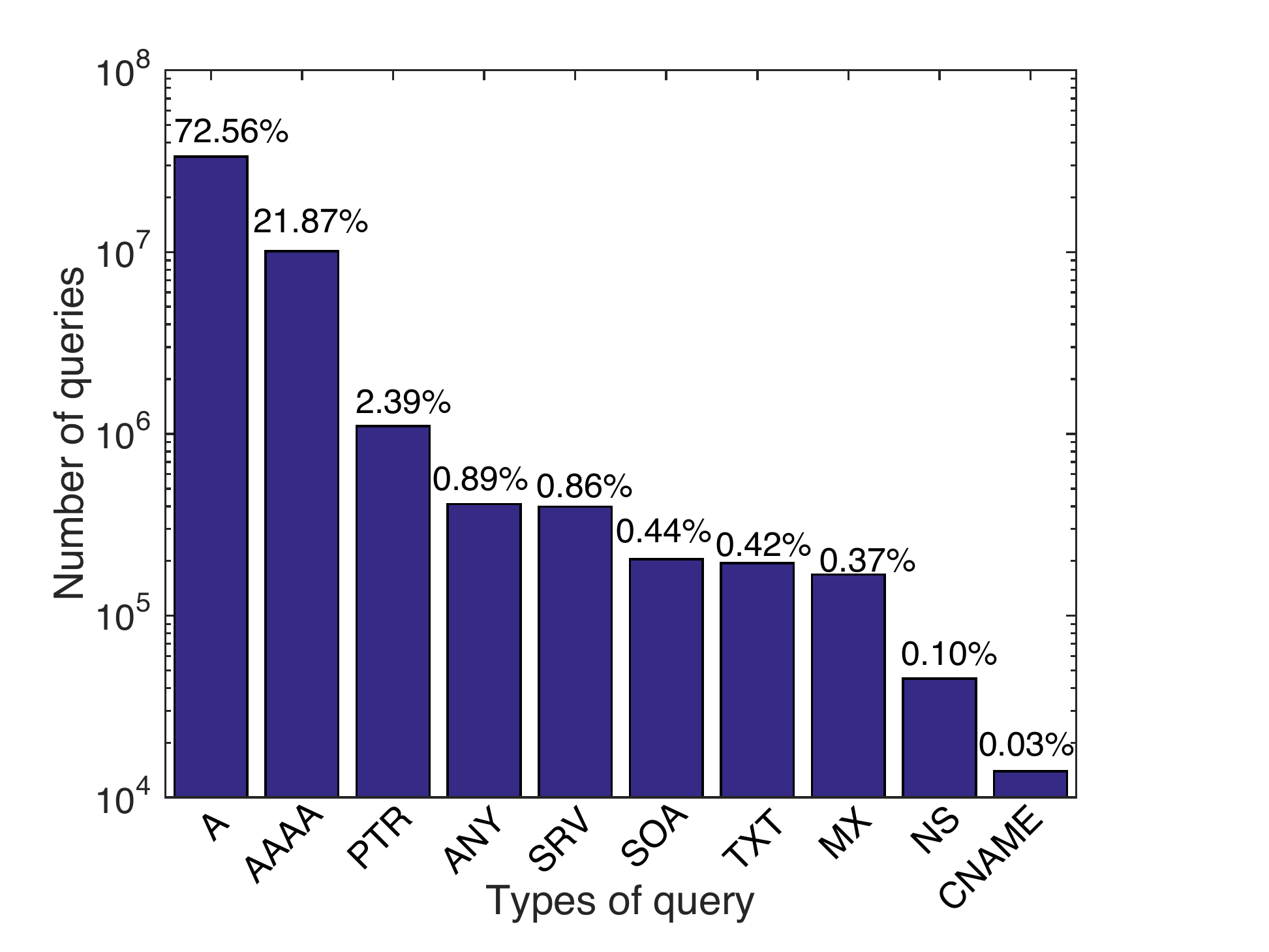}
\label{fig:Qtype-Histogram}}
\subfloat[Top 50 frequently queried domains \protect \\   (include $21.39\%$ of all queries)]{\includegraphics[width=1.7in]{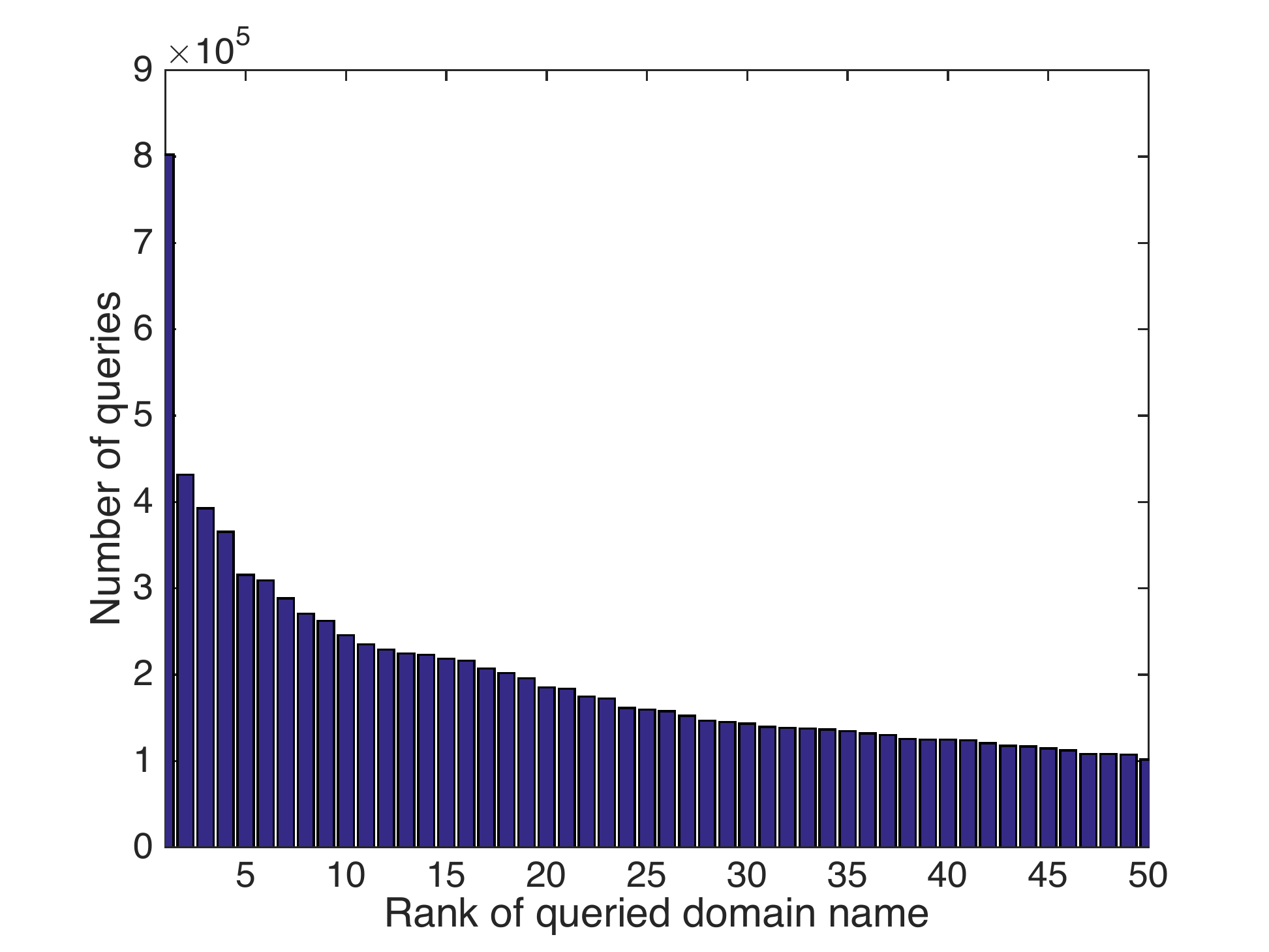}
\label{fig:Qname-Histogram}
}
\hfil
\subfloat[Top 50 frequent source IPs  \protect \\   (include $4.56\%$ of all queries)]{\includegraphics[width=1.7in]{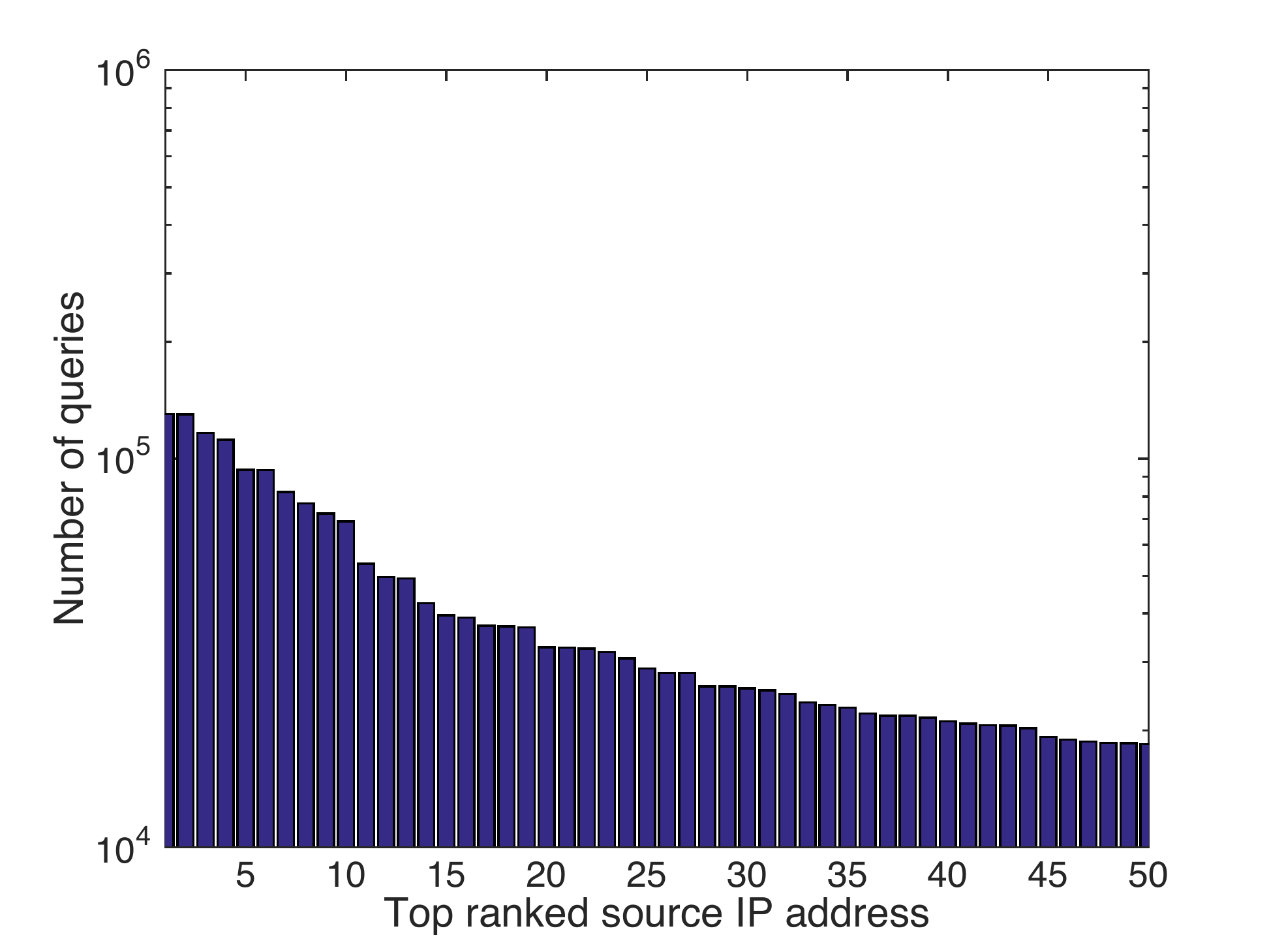}
\label{fig:Src-Histogram}
}
\subfloat[Top 50 frequent destination IPs \protect \\ (include $99.77\%$ of all queries)]{\includegraphics[width=1.7in]{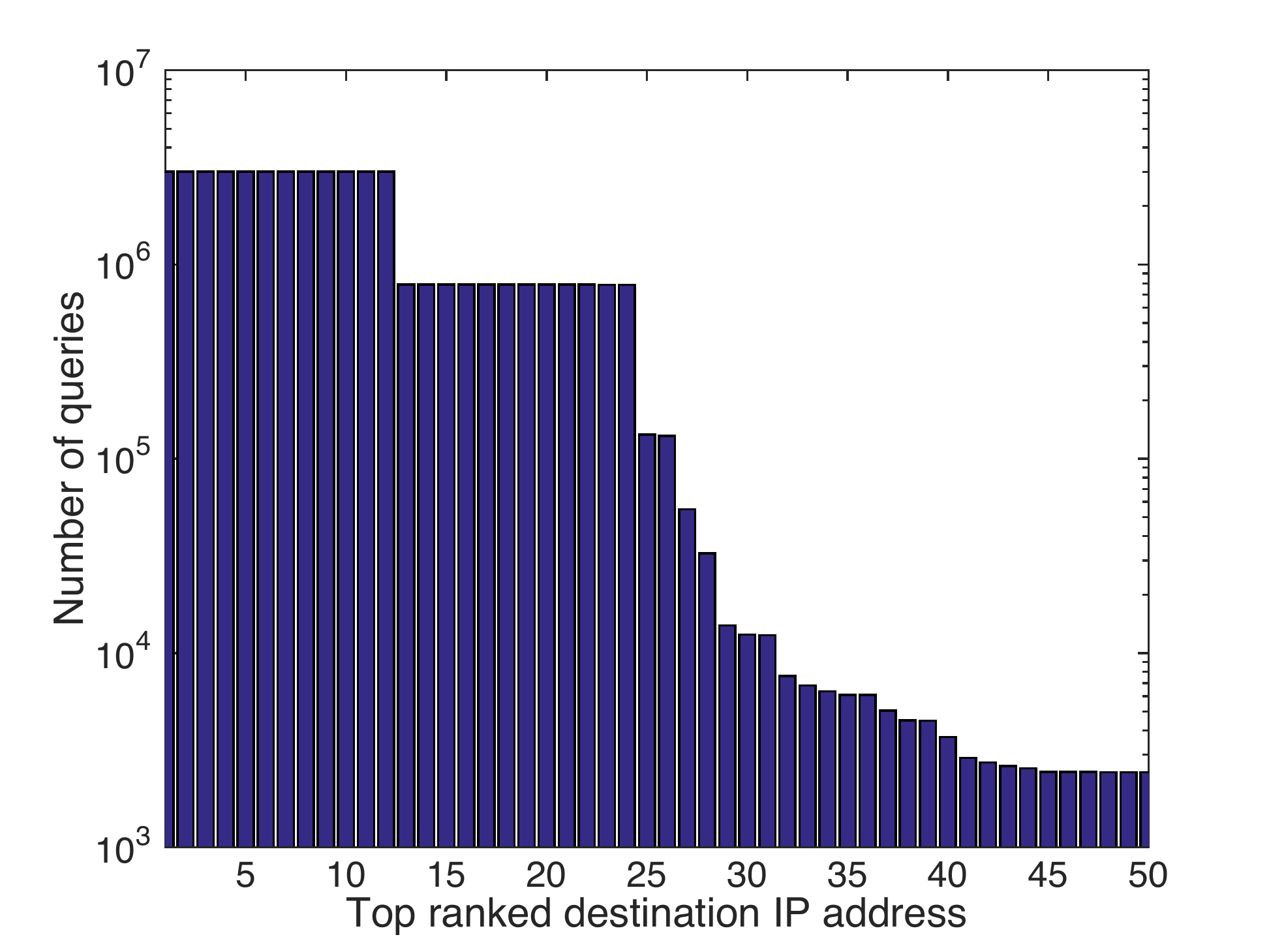}
\label{fig:Dst-Histogram}
}
\caption{Histograms of several basic fields of DNS queries.}
\label{fig:Overall-Histogram}
\end{figure}

\figref{Overall-Histogram} shows histograms of some basic information about our DNS query data. For the histogram of query types, the figure only shows the top 10 most frequent query types. More than $94\%$ of the queries are for IPv4 and IPv6 address records, and $2.39\%$ are domain name pointer queries.  Histograms demonstrate the distribution of network traffic, and the anomalous change of histogram can indicate traffic anomalies \cite{LakCro2005}. For example, during a DDOS attack, the distribution of source IP addresses may be more disperse than normal distributions since during such attack large amount of traffic are initiated from a large number of unique IP addresses \cite{LakCro2005}.

\begin{figure}[!t]
\centering
\subfloat[Norm]{\includegraphics[width=1.16in]{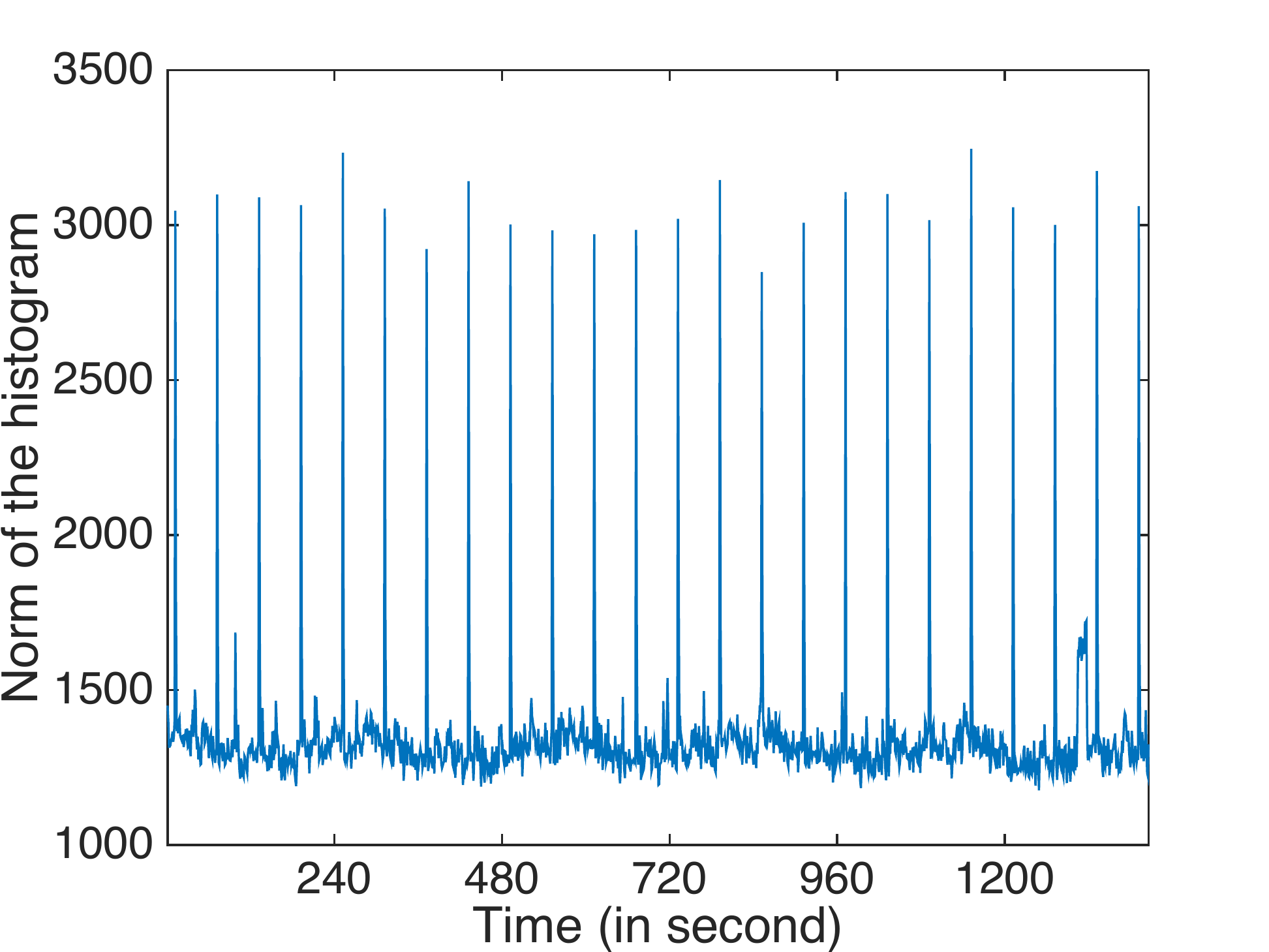}
\label{fig:Histogram-Norm}}
\subfloat[Entropy]{\includegraphics[width=1.16in]{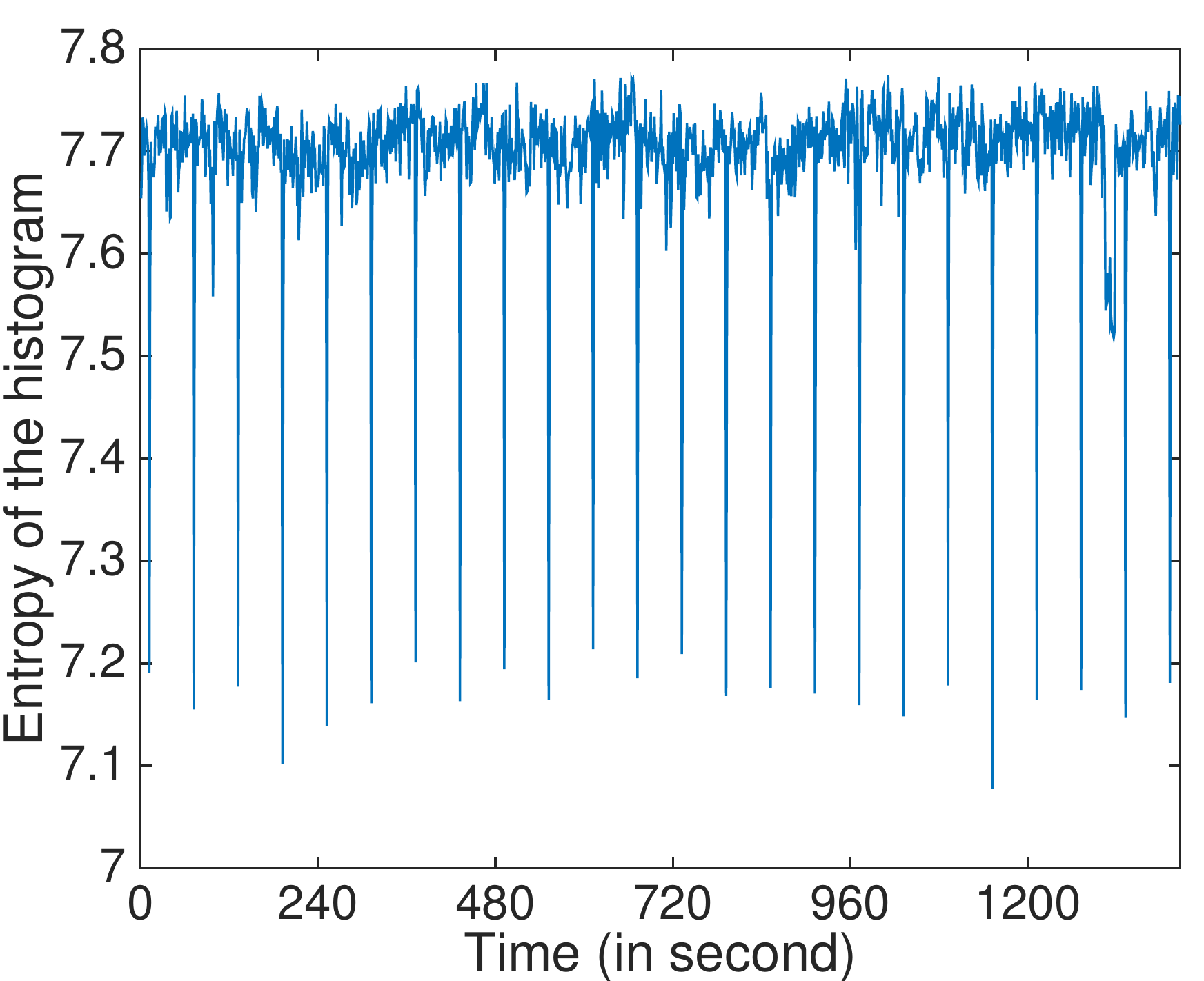}
\label{fig:Histogram-Entropy}}
\subfloat[Fraction]{\includegraphics[width=1.16in]{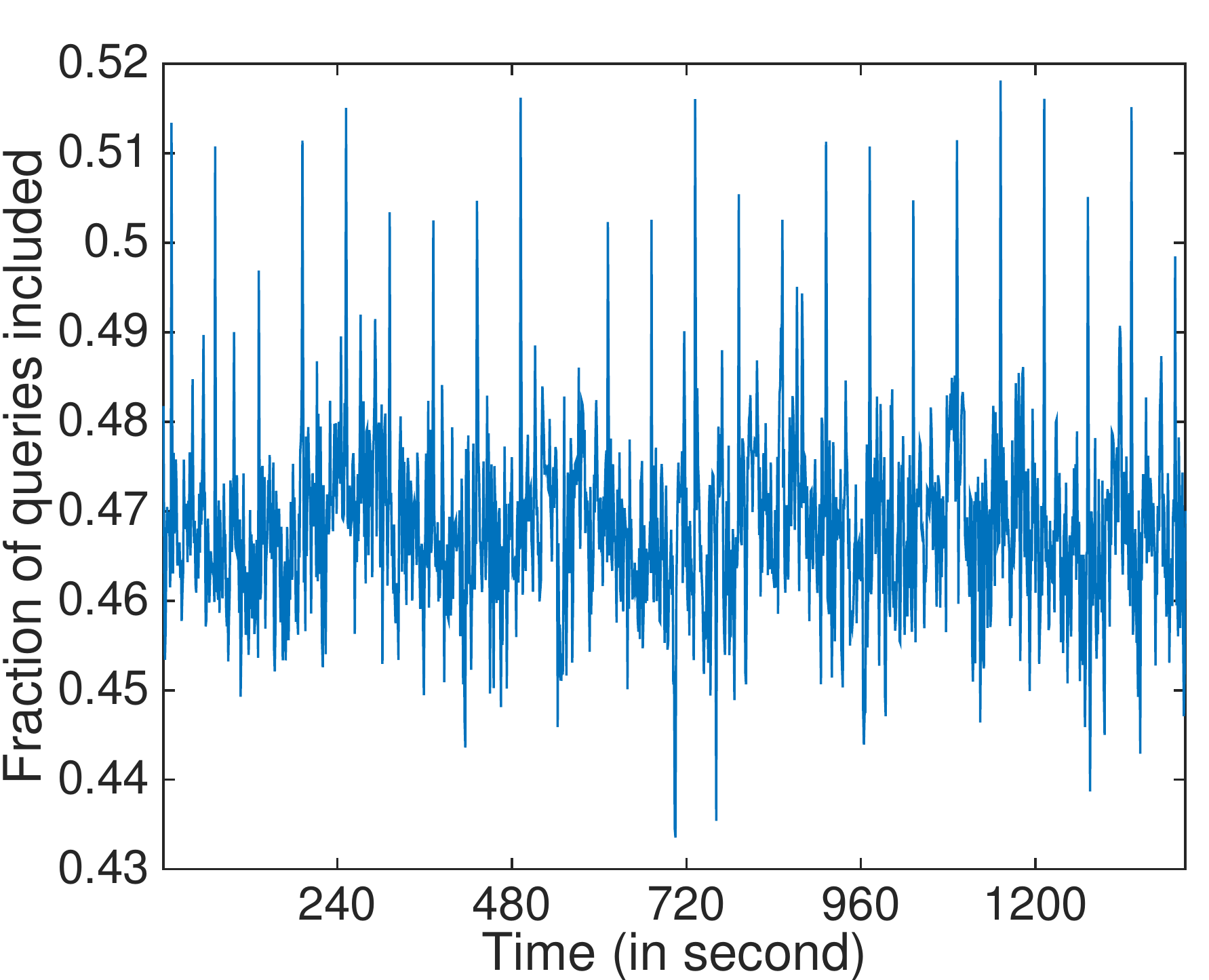}
\label{fig:Histogram-Fraction}}
\caption{Norm and entropy of the histogram, and the fraction of the queries included in the histogram.}
\label{fig:Histogram-NormAndEntropy}
\end{figure}

We next discuss how to aggregate the raw DNS data. We first group the DNS packets based on the query timestamp into 1406 bins, one bin for each second in the dataset. In each time bin, we compute a histogram indicating the distribution of traffic volume over all possible domain names queried in that interval. For each histogram, we only kept the frequencies of the top 300 most frequently queried domain names. \figref{Histogram-Norm} and \figref{Histogram-Entropy} shows the norms and entropies of the histogram versus various time bins respectively, and \figref{Histogram-Fraction} demonstrates the fraction of queries included in our truncated histograms over the total number of queries per second.  When a histogram is more uniformly distributed, it has a corresponding higher entropy, and vice-versa. In \figref{Histogram-NormAndEntropy}, there is a periodic sharp increase of the histogram norm and a decrease of the entropy both with a period of 60 seconds.

\begin{figure}[!t]
\centering
\includegraphics[width=2.4in]{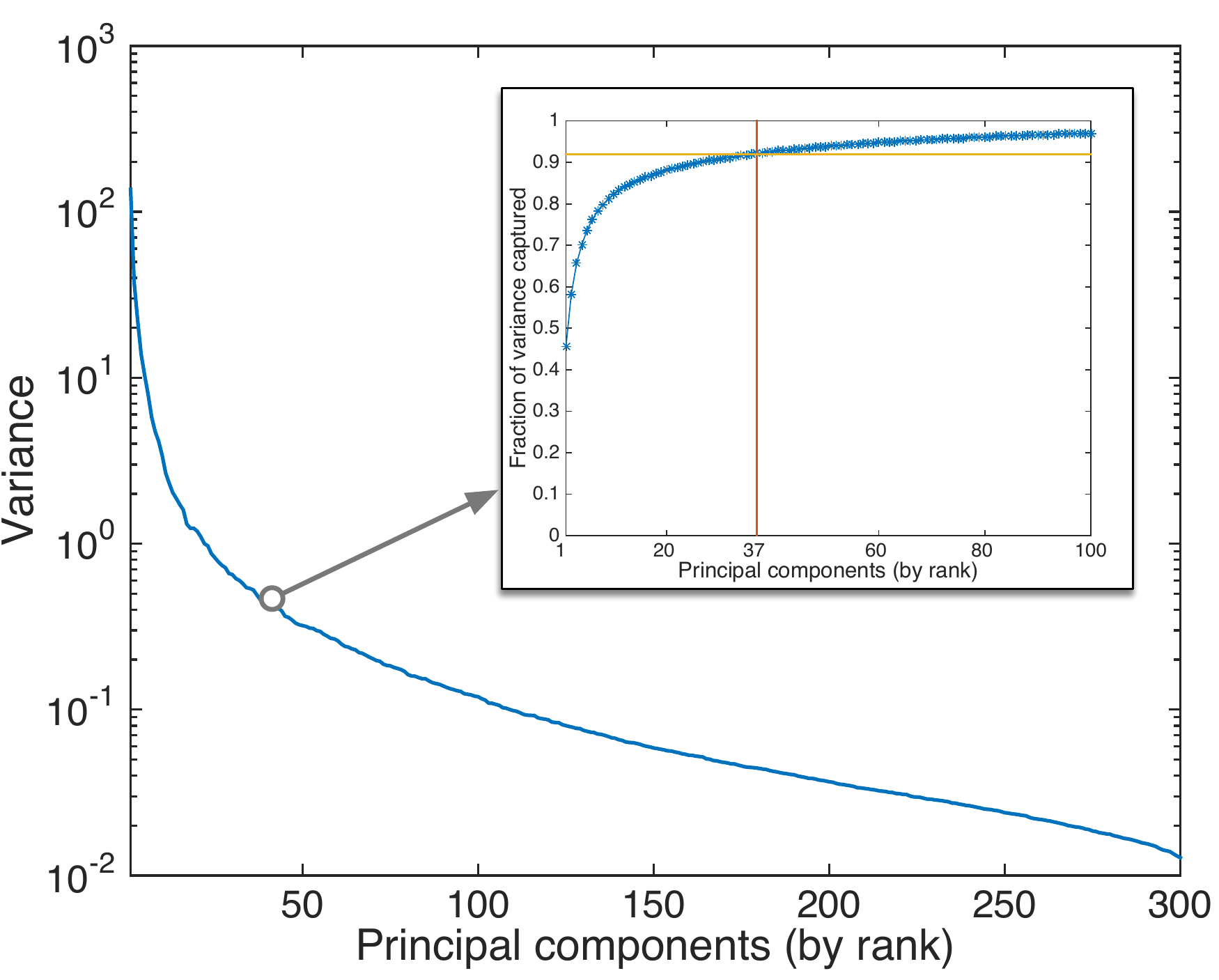}
\caption{Scree plot of DNS dataset.}
\label{fig:Scree-Plot}
\end{figure}

One key issue with PCA is to determine the dimension of the principal subspace.  Setting the size of the principal subspace either too high or too low may degrade the anomaly detection accuracy. One general approach for determining the dimension is to set a variance threshold so that a majority of the variance is captured in the principal subspace. Typical choices of this threshold are over $90\%$. \figref{Scree-Plot} shows the scree plot of the dataset used in this paper on log scale (note: we normalized the dataset column-wise before performing PCA on it).  It shows the change of variance captured by the PCs ordered by rank.  The mini-plot inside of the scree plot shows the cumulative fraction of the variance captured by the PCs. We can see although the dimension of our dataset is 300, it has a very low intrinsic dimensionality in that we only need $9\%$ of the PCs to capture more than $90\%$ of the variance. We set our variance threshold to capture $92\%$ of the variance, and the corresponding principal subspace dimension is 37. 

\subsection{Experiments}
Our dataset is a $1406\times 300$ matrix, each row of which is a histogram vector of length $300$, aggregated using the method introduced in \secref{Dataset}. First, we measure the error of the approximate dominant PCs found by distributed PCA algorithms with regards to the true PCs found by centralized PCA.  Second, we test the accuracy of distributed PCA anomaly detection methods compared with the centralized method. 

\subsubsection{Accuracy of subspace estimation}
We use geodesic distance (GD) as a metric to evaluate the difference between the estimated principal subspace and the true one. As a widely-used distance metric for quantifying the distance between two linear subspaces, GD is derived based on the principal angles \cite{HamJih2008}, which are the angles between two subspaces spanned by orthonormal matrices $\mathbf{\hat{V}}$ and $\mathbf{{V}}$:
\begin{equation}
\begin{aligned}
\cos \theta_i = & \max_{u_i\in\mbox{span}(\hat{\mathbf{V}})}\max_{v_i\in\mbox{span}(\mathbf{{V}})} u_i^{\intercal}v_i\,\,\, \mbox{      subject to } \\
& ||u_i||=||v_i||=1 \\
& u_i^{\intercal} u_j = 0, v_i^{\intercal}v_j = 0\,\,\,\, \forall j=\{1,...,i-1\} .
\end{aligned}
\end{equation}
Geodesic distance is defined as $d(\mathbf{{V}},\hat{\mathbf{V}})=\sqrt{\sum_{i=1}^k\theta_i^2 }$.
Cosines of the principal angles are equal to the singular values of $\hat{\mathbf{V}}^{\intercal}\mathbf{{V}}$ \cite{HamJih2008}. Thus we can use the singular values to find the corresponding principal angles.

There are two design parameters for distributed PCA: $i)$ $r$, the number of local PCs sent to the DFC, and $ii)$ $s$, the number of partitions (i.e.,\ local monitors).  We want to investigate how $r$ and $s$ will affect the estimation accuracy. For the horizontal partitioning experiment, we first horizontally partitioned the original dataset into $s$ sub-matrices, each of which is $[1406/s]\times 300$ matrix ($[\cdot]$ denotes the nearest integer) except for the last matrix which may have slightly more or fewer than $[1406/s]$ rows. For the vertical partitioning experiment, we partitioned the original dataset into $s$ parts, each of which is a $1406\times [300/s]$ matrix. Following the distributed PCA methods for horizontal and vertical partitioned matrices, we can approximate the true principal subspace, and then compute the GD between the approximated principal subspace and the true one. \figref{Horizontal-Vertical-GD} shows our results when $s$ (i.e., the number of partitions) is $2$ or $4$, and shows the change of GD with regard to $r$ (i.e., the number of local PCs). As observed in \figref{Horizontal-Vertical-GD}, the GD decreases as we send more local PCs to the DFC. This decreasing trend is very sharp at the start, but it slows down when $r$ reaches a certain level. Notice that there are two curves in each figure: one is for GD, and the other one is for normalized communication cost, which will be covered in the following section. 

\subsubsection{Communication cost of distributed PCA}
The normalized communication cost is defined as the communication cost of using distributed PCA (the number of values sent), normalized by the number of values in the overall matrix $m n$, as the latter is proportional to the cost of simply sending the matrix to the DFC and doing centralized PCA. For horizontally or vertically partitioned datasets, the respective costs are
\begin{eqnarray}
\label{eq:Horizontal-Normalized-Comm}
c_{\rm hor} &=& \frac{\sum_{i=1}^s(r+nr)}{mn}=\frac{sr(n+1)}{mn} \\
\label{eq:Vertical-Normalized-Comm}
c_{\rm ver} &=& \frac{\sum_{i=1}^s(mr+n_i r)}{mn}=\frac{sr(m+n/s)}{mn}.
\end{eqnarray}
It is natural to consider the limiting case of normalized communication cost in \eqref{Horizontal-Normalized-Comm} and \eqref{Vertical-Normalized-Comm}
as $n$ grows large with $s, m, r$ fixed:
\begin{equation}
\lim_{n\rightarrow \infty} c_{\rm hor} = \frac{sr}{m}  \qquad\text{and}\qquad  
\lim_{n\rightarrow \infty} c_{\rm ver} = \frac{r}{m}.
\end{equation}
Likewise, the limits as $m$ grows large for fixed $s, n, r$ are:
\begin{equation}
\lim_{m\rightarrow \infty} c_{\rm hor} = 0  \qquad\text{and}\qquad  
\lim_{m\rightarrow \infty} c_{\rm ver} = \frac{sr}{n}.
\end{equation}
We can see that for a extremely high-dimensional dataset, the communication cost of vertically partitioning will be lower than horizontally partitioning. However, when the number of samples grows large, the communication cost for horizontal partitioning is negligible compared with sending the whole dataset, and it is higher than vertical partitioning.

\begin{figure}[!t]
\centering
\subfloat[Horizontal partitioning with s=2]{\includegraphics[width=1.7in]{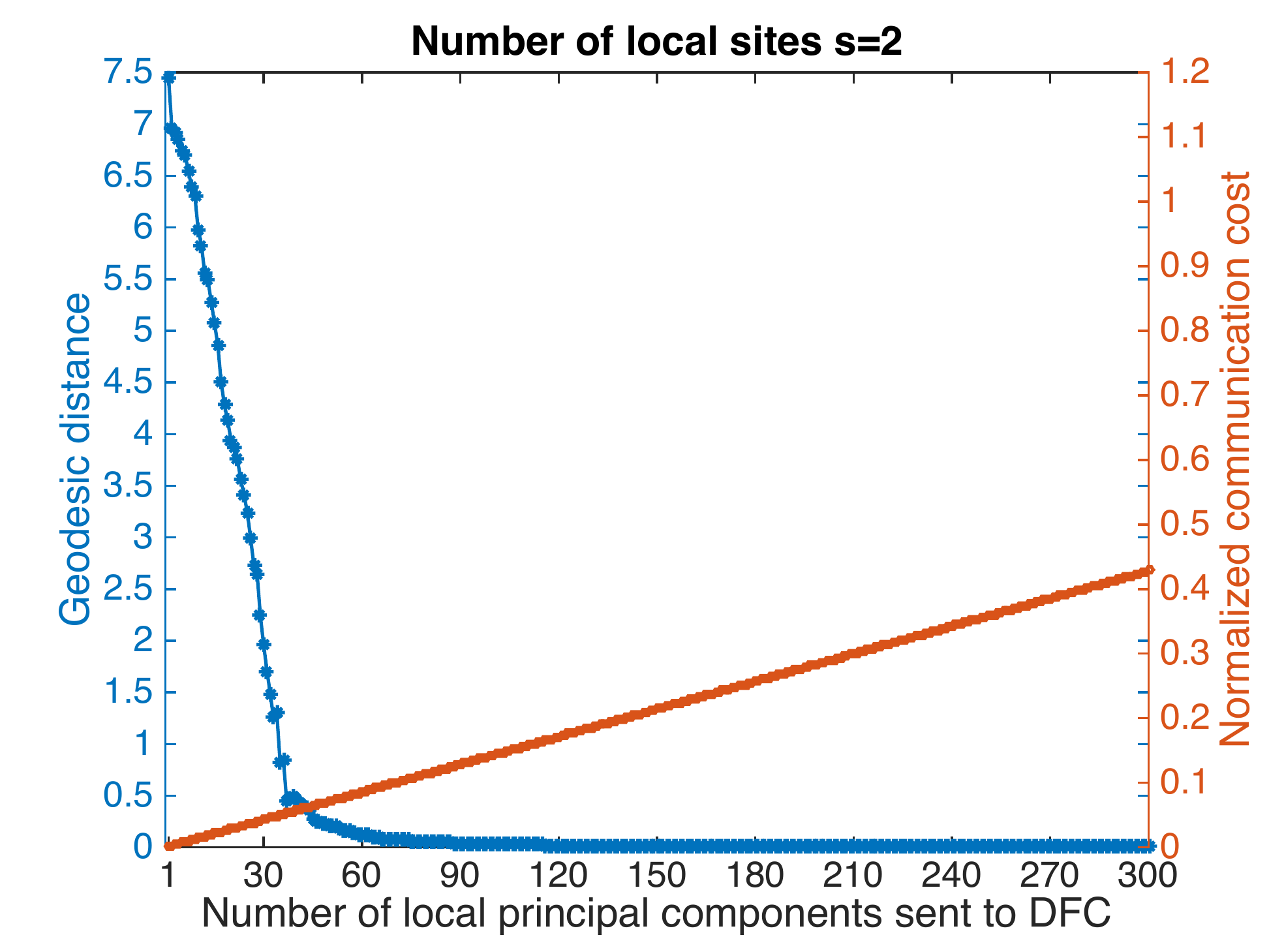}
\label{fig:Horizontal-GD-s2}}
\subfloat[Horizontal partitioning with s=4]{\includegraphics[width=1.7in]{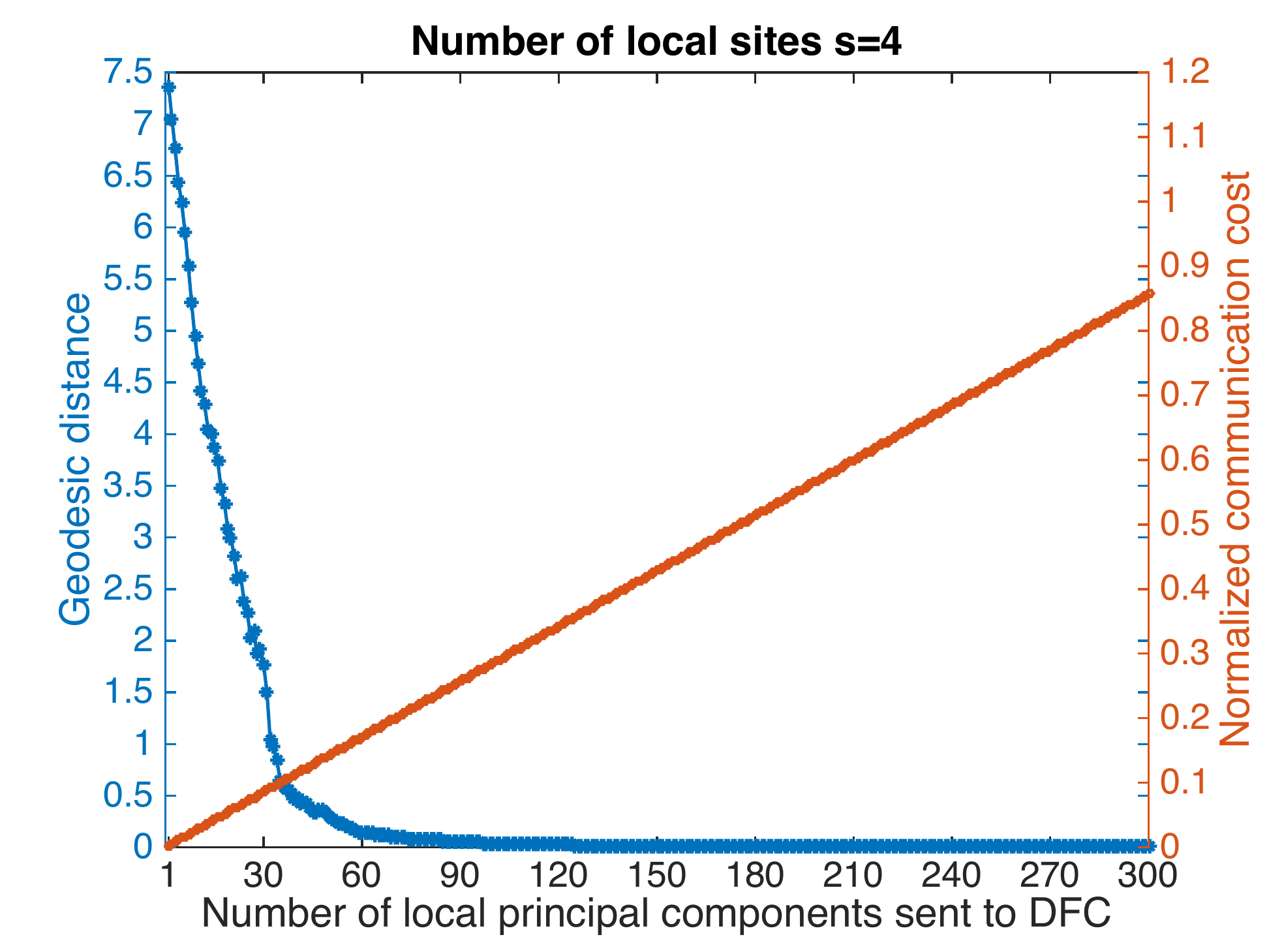}
\label{fig:Horizontal-GD-s4}}
\hfil
\subfloat[Vertical partitioning with s=2]{\includegraphics[width=1.7in]{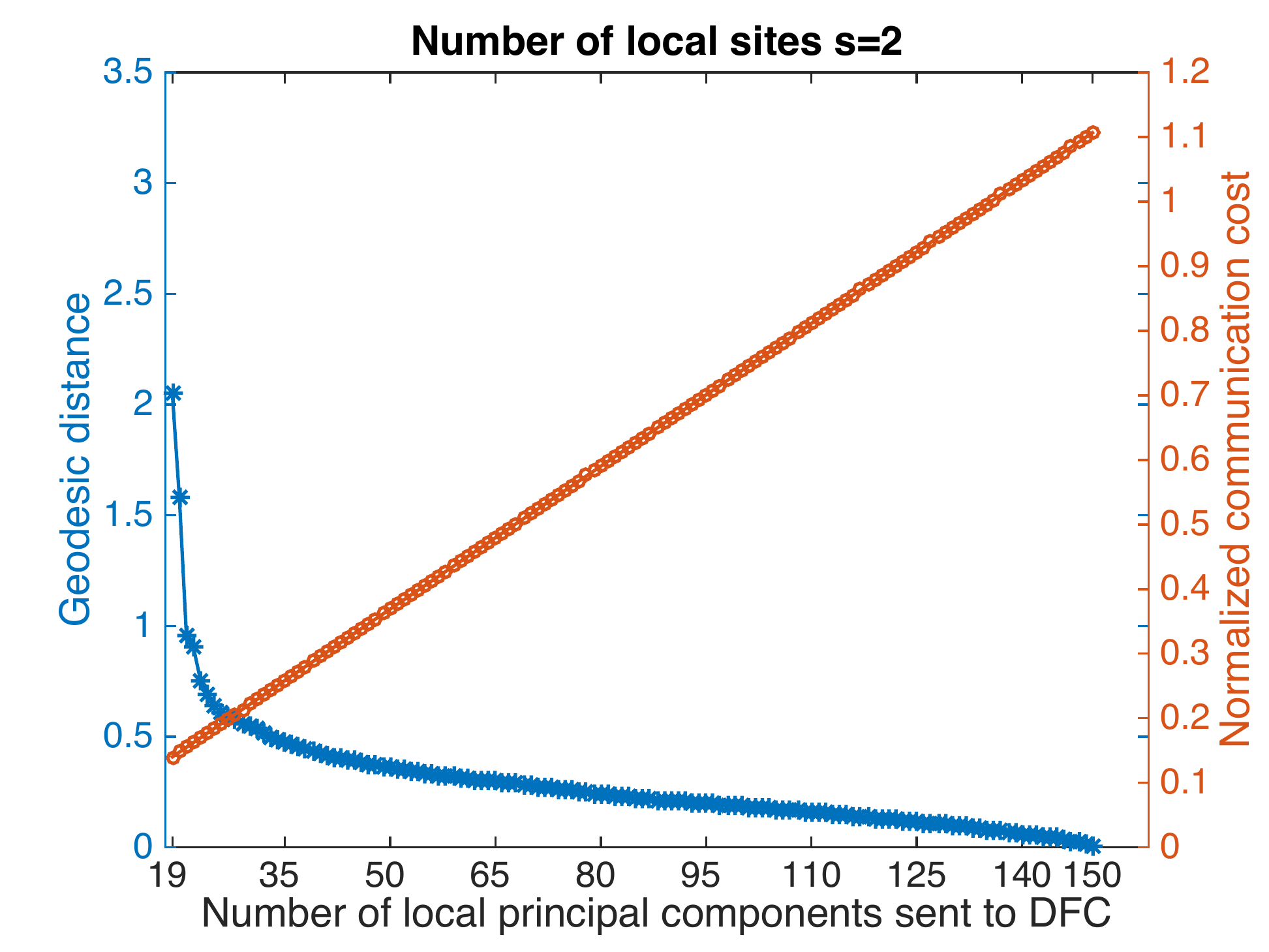}
\label{fig:Vertical-GD-s2}}
\subfloat[Vertical partitioning with s=4]{\includegraphics[width=1.7in]{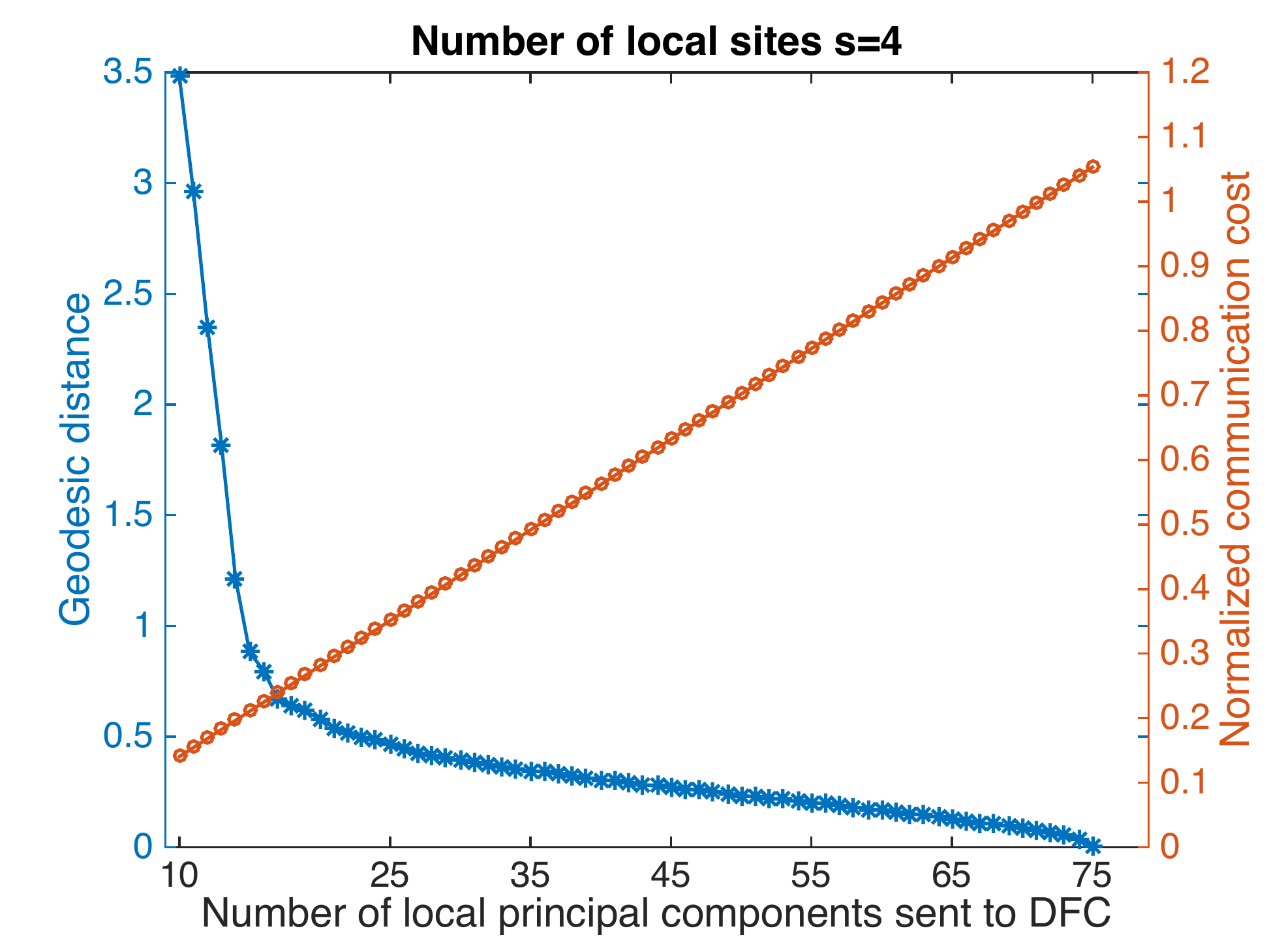}
\label{fig:Vertical-GD-s4}}
\caption{Comparing the GD between the approximate principal subspace and the true principal subspace in horizontal partitioning case and vertical partitioning case.}\label{fig:Horizontal-Vertical-GD}
\end{figure}

Observed in \figref{Horizontal-GD-s2}, distributed PCA of horizontally partitioned case only uses $10\%$ of the centralized approach's communication cost to achieve a GD as low as 0.0774 when $s=2$.  However, to reach the same accuracy when there are $s=4$ local monitors, we have to spend $22\%$ of the communication cost of sending the whole dataset.  Comparing the upper two plots with the lower two in \figref{Horizontal-Vertical-GD}, we can see the normalized communication cost of the vertical partitioning approach is typically higher than the horizontal partitioning case. For vertical partitioning case, the communication cost increases fast as we increase the number of local PCs sent to the DFC (i.e., $r$). Also seen in \figref{Vertical-GD-s2} and \figref{Vertical-GD-s4}, the communication cost may grow close to or even above one for a large $r$. This is because we include information about the left singular vectors in the local datasets sent to the DFC. 

\begin{figure}[!t]
\centering
\subfloat[Minimum $c_{\rm ver}$ / $c_{\rm hor}$ for GD $\leq d^*$.]{\includegraphics[width=2.4in]{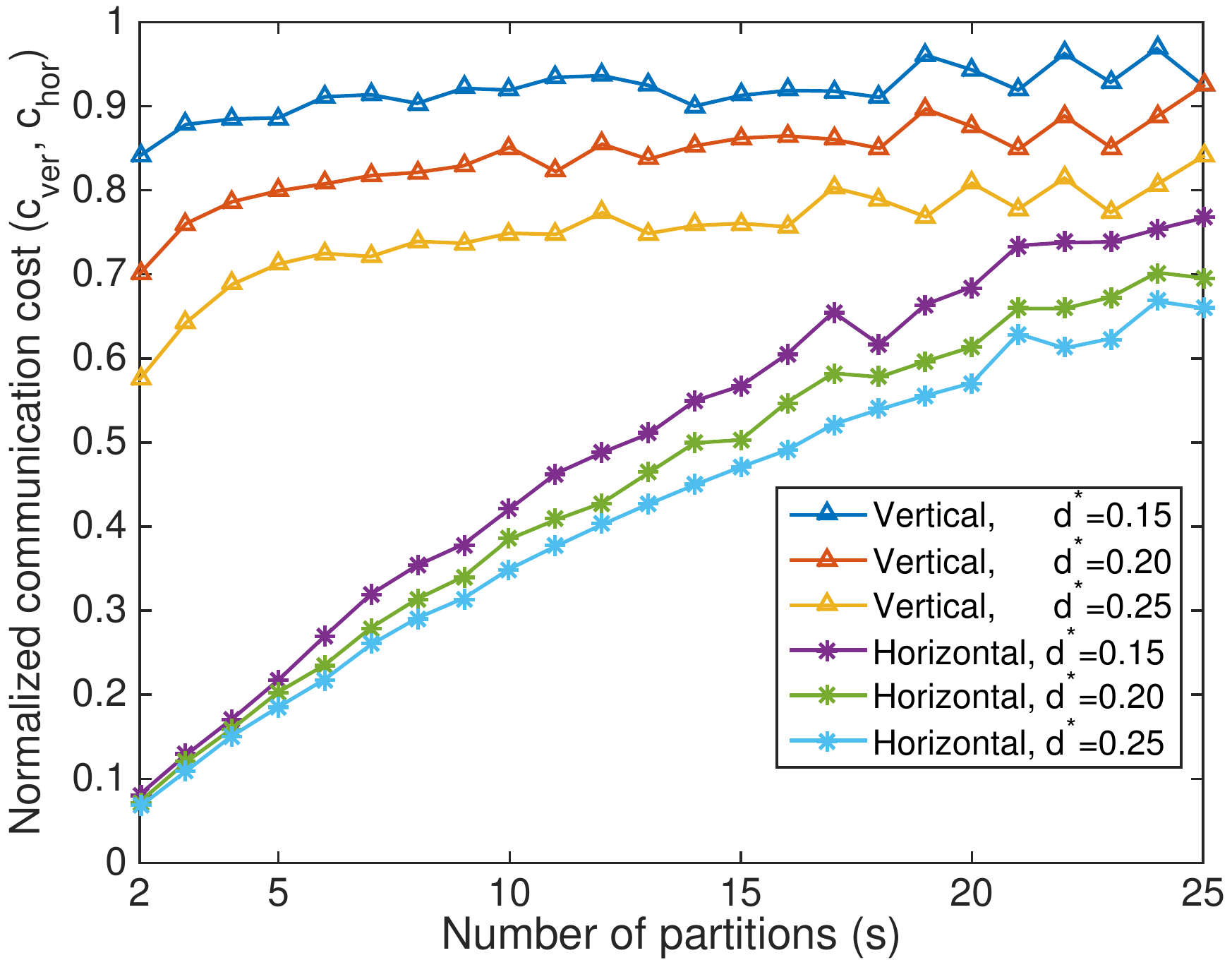}
\label{fig:Optimal-Comm}
}
\hfil
\subfloat[Minimum $r$ for GD $\leq d^*$.]{\includegraphics[width=2.4in]{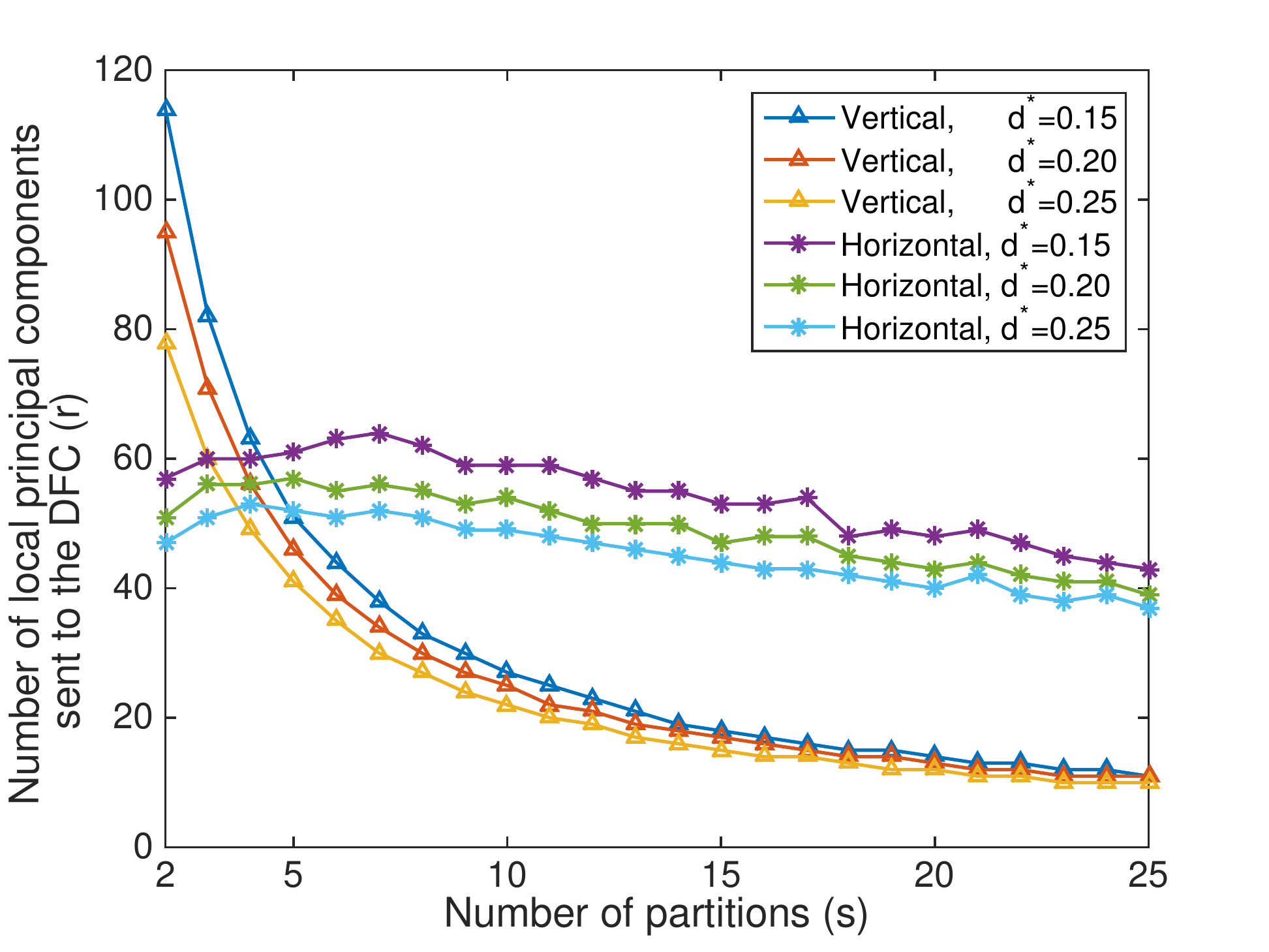}
\label{fig:Optimal-R}}
\caption{$c_{\rm hor}$ (or $c_{\rm ver}$) and $r$ as a function of $s$ for various $d^*$.}
\label{fig:Optimal-R-Comm}
\end{figure}

Next we varied the number of partitions $s$ (i.e., the number of local monitors) from 2 to 25, and chose a maximum tolerable GD threshold $d^*$. For given $s$ and $d^*$, we then find the minimum number of local PCs $r^*(s, d^*)$ such that its corresponding GD is less than or equal to $d^*$. For each $r^*(s,d^*)$, we computed the normalized communication cost as \eqref{Horizontal-Normalized-Comm} and \eqref{Vertical-Normalized-Comm}. According to \figref{Optimal-Comm}, the overall trend of communication cost is increasing with $s$, and a higher communication cost is required for achieving a lower $d^*$. It's also clear that the required $c_{\rm ver}$ is typically higher than $c_{\rm hor}$, indicating that the vertical case requires more communication overhead than horizontal case to achieve the same $d^*$. \figref{Optimal-R} shows the relationship between $r^*(s, d^*)$ versus $s$. We can see that for the vertical partitioning case, as the number of partitions increases, the number of required local PCs sent to the DFC quickly decreases. Compared with horizontal partitioning, the number of required local PCs under vertical partitioning is more sensitive to $s$.

\subsubsection{Accuracy of anomaly detection}

\begin{figure}[!t]
\centering
\subfloat[Horizontal partitioning and \protect \\ ground truth threshold is $1\%$ ]{\includegraphics[width=1.7in]{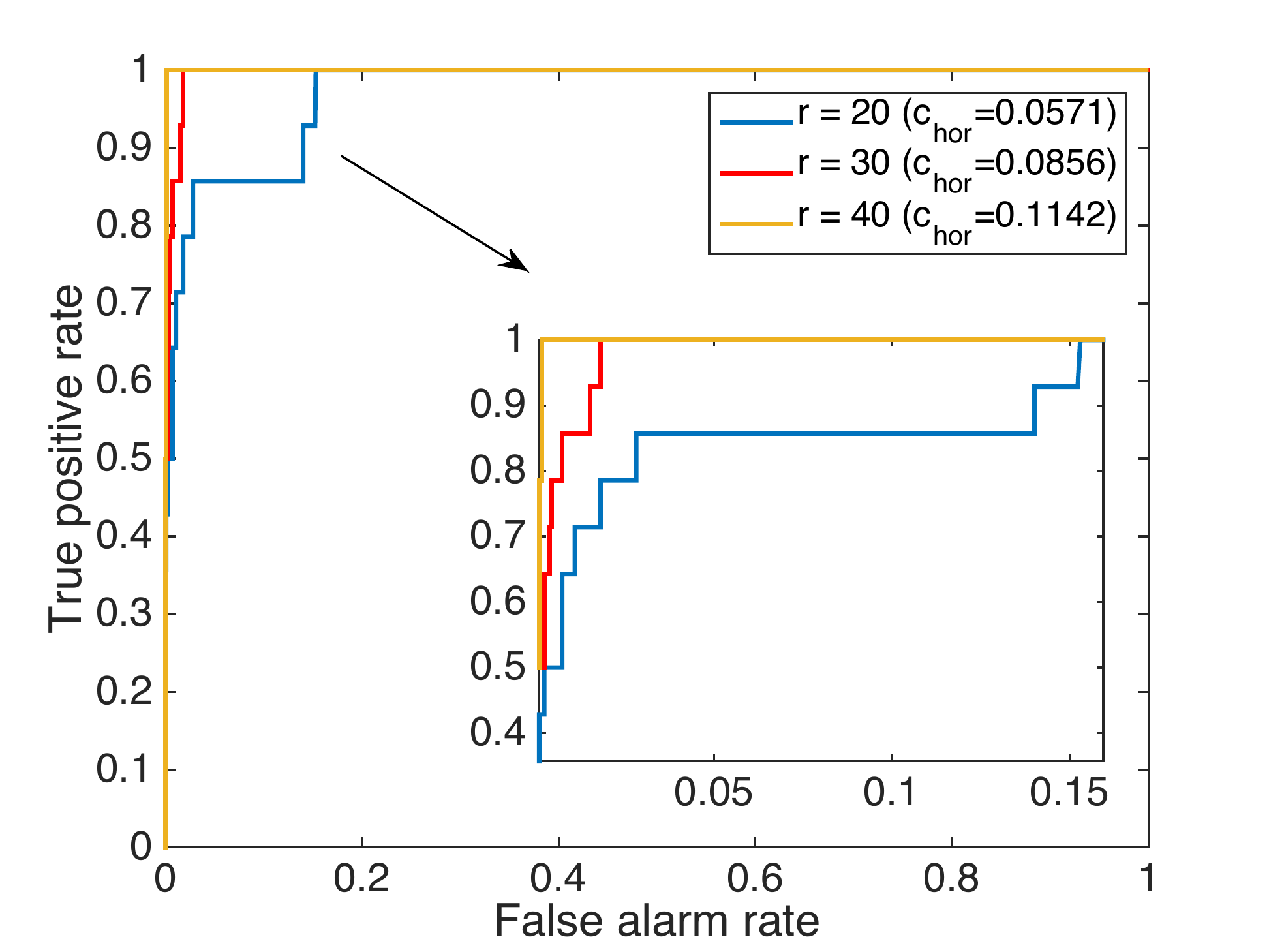}
\label{fig:ROC-Horizontal-001}}
\subfloat[Vertical partitioning and ground truth threshold is $1\%$]{\includegraphics[width=1.7in]{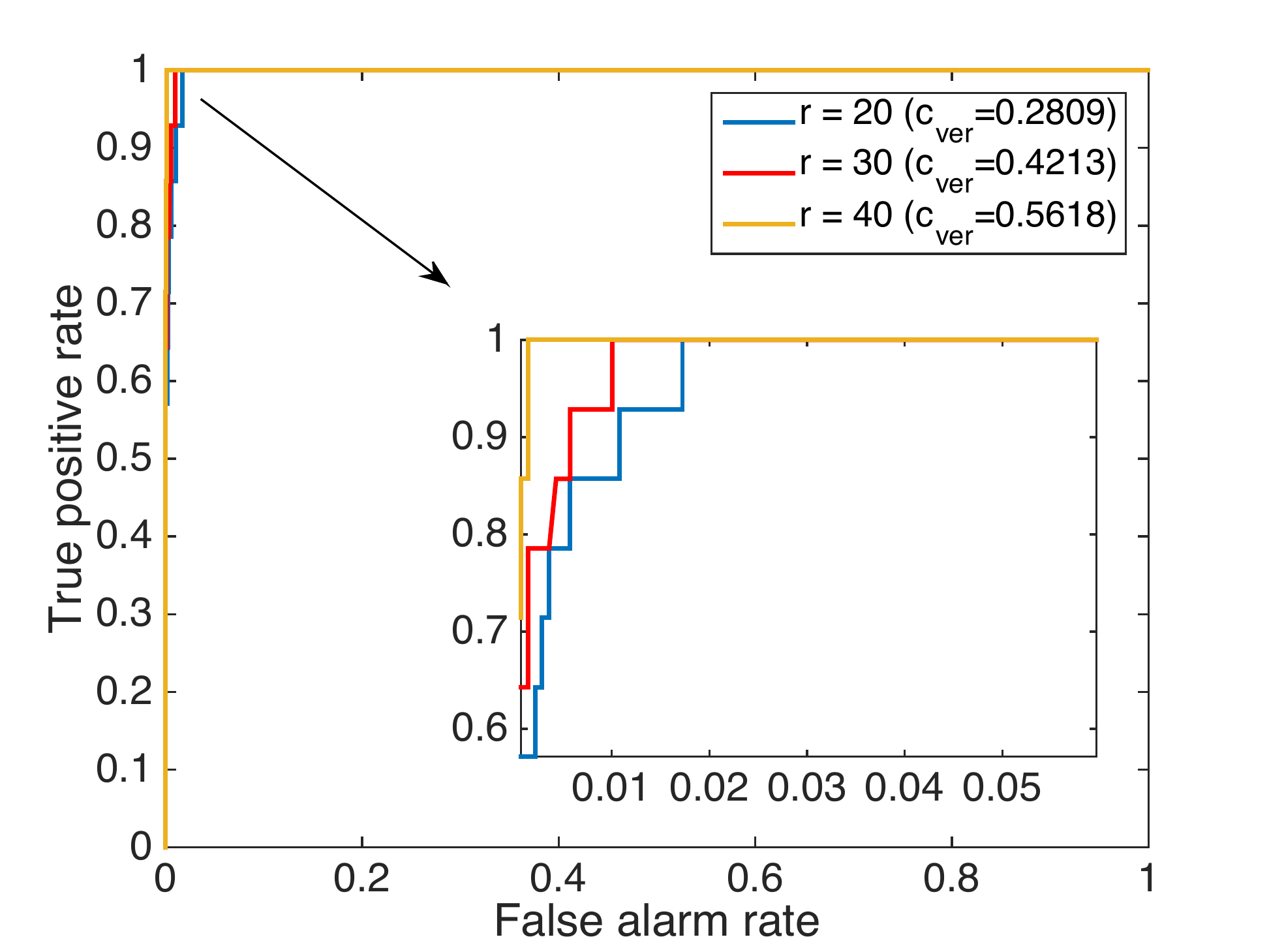}
\label{fig:ROC-Vertical-001}}
\hfil
\subfloat[Horizontal partitioning and\protect \\  ground truth threshold is $5\%$]{\includegraphics[width=1.7in]{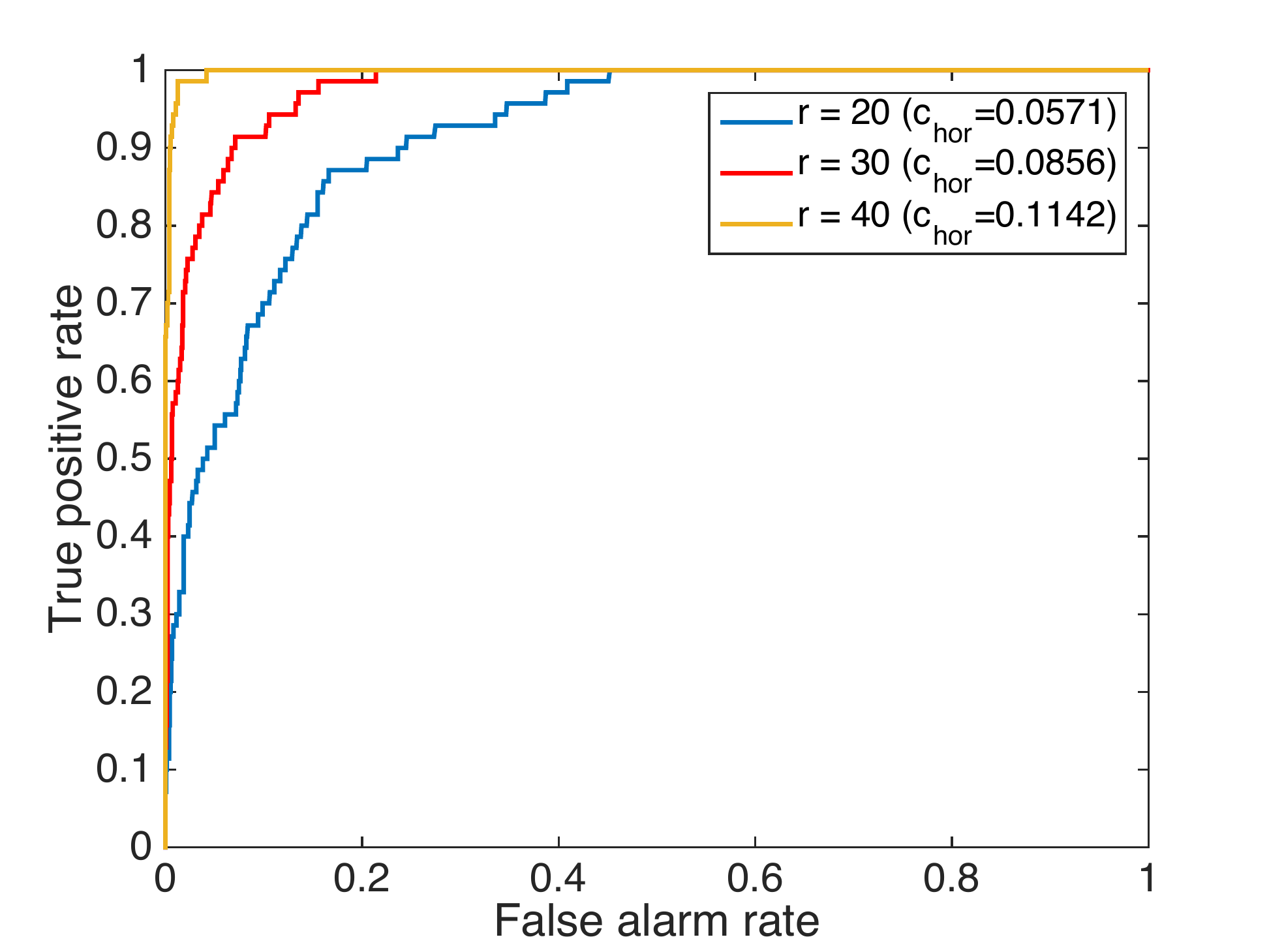}
\label{fig:ROC-Horizontal-005}}
\subfloat[Vertical partitioning and ground truth threshold is $5\%$]{\includegraphics[width=1.7in]{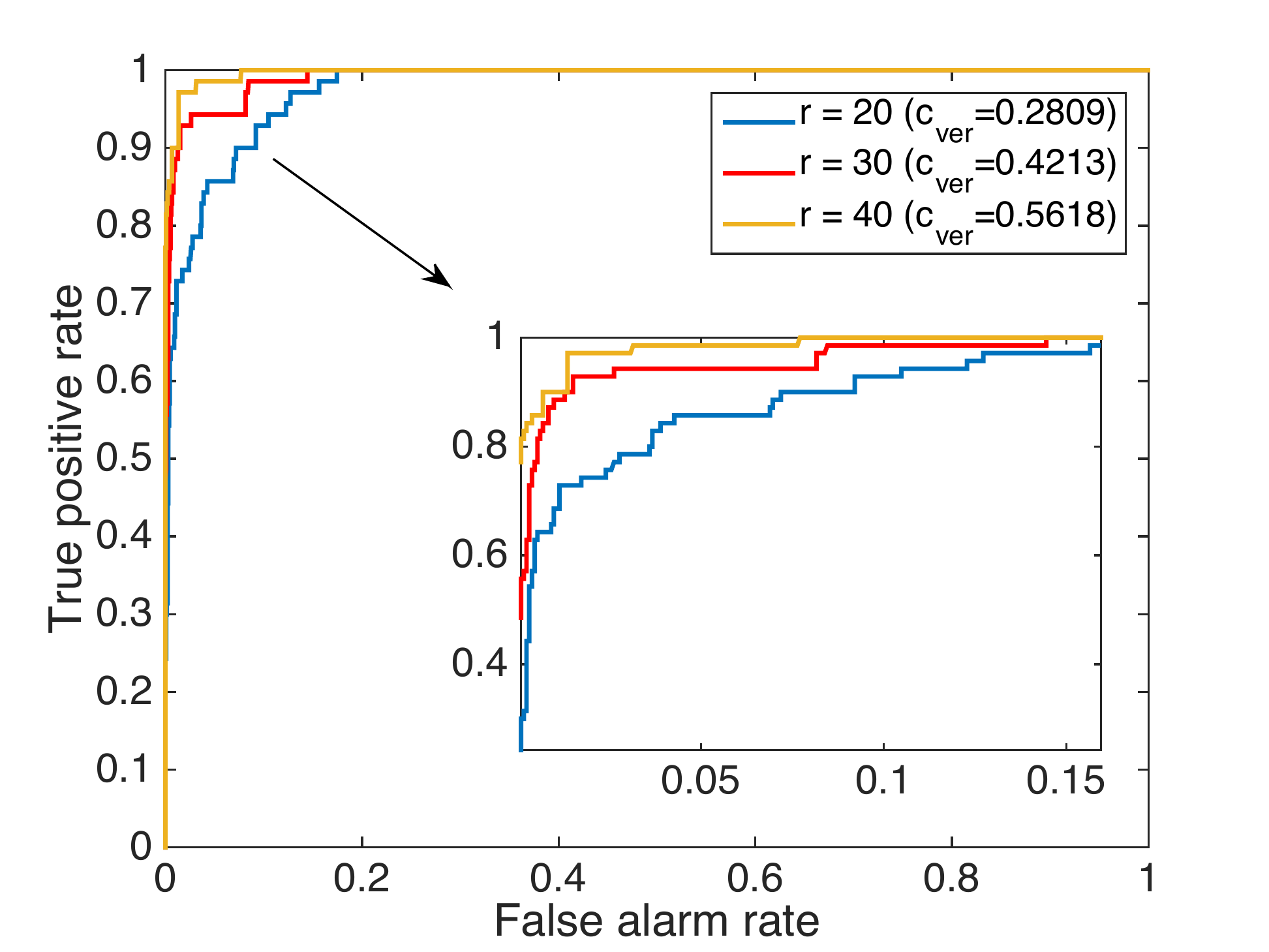}
\label{fig:ROC-Vertical-005}}
\hfil
\subfloat[Horizontal partitioning and\protect \\  ground truth threshold is $10\%$]{\includegraphics[width=1.7in]{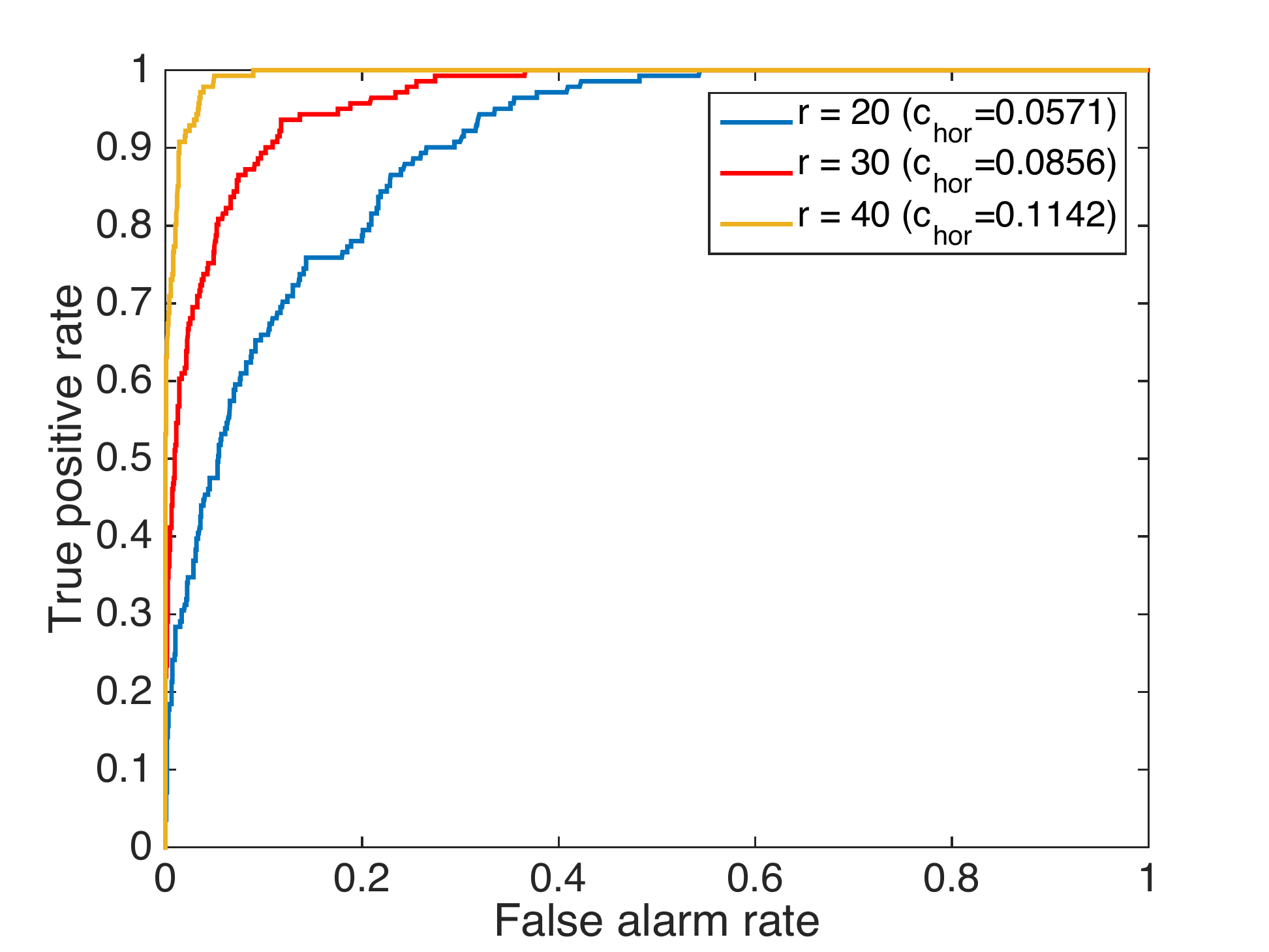}
\label{fig:ROC-Horizontal-01}}
\subfloat[Vertical partitioning and ground truth threshold is $10\%$]{\includegraphics[width=1.7in]{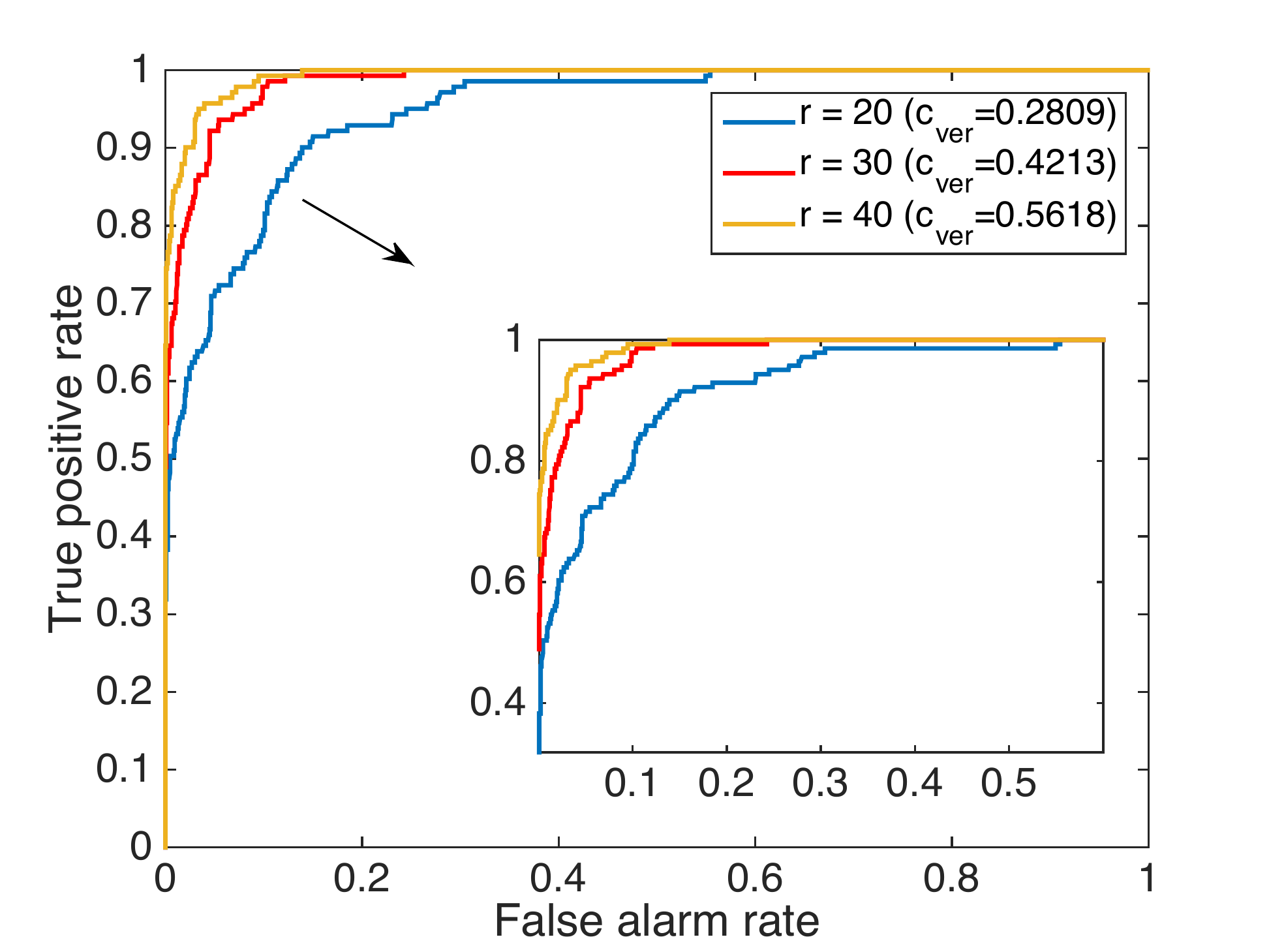}
\label{fig:ROC-Vertical-01}}
\caption{The ROC curves of distributed PCA algorithms.}
\label{fig:Horizontal-Vertical-ROC}
\end{figure}

We tested the performance of the two distributed PCA based anomaly detection methods on the same DNS dataset, with the dataset partitioned into 4 parts. To set up \emph{ground truth}, we used centralized PCA-based subspace method to label anomalies and treat them as ``true'' anomalies. We chose three different detection thresholds to include the top $1\%$, $5\%$, and $10\%$ samples with the largest squared residual norm as three ground truth sets. Next, using each ground truth set, we implemented the distributed PCA anomaly detection algorithms of vertical partitioning and horizontal partitioning, and created three receiver operating characteristic (ROC) curves for each case with three different $r$ values, see \figref{Horizontal-Vertical-ROC}. On ROC curve, the x-axis and y-axis shows the false alarm rate (FAR) and the true positive rate (TPR) respectively. As we send more information to the DFC, the detection performance becomes better. \figref{ROC-Horizontal-001} and \figref{ROC-Vertical-001} show that it is enough to send only the top $20$ local PCs' information in order to reach a good detection accuracy with a very low FAR, and the corresponding normalized communication costs are $0.0571$ and $0.2809$ for horizontal partitioning and vertical partitioning respectively. This figure indicates that we can achieve a large saving in communication bandwidth, especially for the horizontal partitioning case, with a extremely small loss of detection accuracy. Equal error rate (ERR) is the rate at which the false positive rate and the FAR are equal or have the minimum distance if equality cannot be achieved. Lower ERR indicates better performance of a classifier. Table \ref{tab:ERR} shows the ERR we computed based on the ROC curves in Fig.\ \ref{fig:Horizontal-Vertical-ROC}. We can see from this table that when $r=20$ and $30$, vertical partitioning outperforms horizontal partitioning in that it has a lower ERR; however when $r$ is large (i.e., $r=40$), horizontal partitioning performs slightly better than vertical partitioning, which is very interesting. Since we only experimented with $s=4$ in this experiment, we need to do more experiments for various choices of $s$ before we can reach a more general conclusion. 

\section{Conclusion}\label{sec:Conclusion}

\begin{table}[!t]
\centering
\begin{tabular}{|c|c|c|c|c|c|c|}
\hline
&\multicolumn{2}{|c|}{$r=20$} &\multicolumn{2}{|c|}{$r=30$} &\multicolumn{2}{|c|}{$r=40$}\\
\hline
\pbox{10cm}{Ground truth\\threshold}&Hor. & Ver. & Hor. & Ver. & Hor. & Ver. \\
\hline
$1\%$&  0.1401    & 0.0172 &  0.0180 &0.0101 & 0.0014  & 0.0014 \\
\hline
$5\%$ &  0.1572   & 0.0921& 0.0861  &0.0569 &  0.0142  & 0.0284 \\
\hline
$10\%$ &  0.2055   &0.1273 &   0.1020        & 0.0640& 0.0356   &0.0419 \\
\hline
\end{tabular}
\caption{Equal error rates of the ROC curves in Fig.\ \ref{fig:Horizontal-Vertical-ROC}.}\label{tab:ERR}
\end{table}

By evaluating the distance between the approximate principal subspace of distributed PCA and the true principal subspace on a real DNS dataset, we analyzed the tradeoff between communication cost and the accuracy of distributed PCA, and the impact of the number of partitions on the communication cost and accuracy. We have shown that distributed PCA algorithms can significantly reduce the communication cost while maintaining a high approximation accuracy.  We also compared the performance of horizontal partitioning distributed PCA and vertical partitioning distributed PCA.  For the dataset used in this paper, the latter one does not have communication bandwidth reduction as large as the former one. We also applied distributed PCA algorithms on anomaly detection and evaluated its detection performance. We have shown that with a small sacrifice on the detection accuracy, we can largely reduce the communication cost, especially for the horizontal partitioning case, compared with the centralized PCA method.

\bibliographystyle{IEEEtran}
\bibliography{./CISS-2016}
\end{document}